# Operationalizing Conflict and Cooperation between Automated Software Agents in Wikipedia: A Replication and Expansion of "Even Good Bots Fight"

R. STUART GEIGER, Berkeley Institute for Data Science, University of California, Berkeley
AARON HALFAKER, Wikimedia Research, Wikimedia Foundation

This paper replicates, extends, and refutes conclusions made in a study published in PLoS ONE ("Even Good Bots Fight"), which claimed to identify substantial levels of conflict between automated software agents (or bots) in Wikipedia using purely quantitative methods. By applying an integrative mixed-methods approach drawing on trace ethnography, we place these alleged cases of bot-bot conflict into context and arrive at a better understanding of these interactions. We found that overwhelmingly, the interactions previously characterized as problematic instances of conflict are typically better characterized as routine, productive, even collaborative work. These results challenge past work and show the importance of qualitative/quantitative collaboration. In our paper, we present quantitative metrics and qualitative heuristics for operationalizing bot-bot conflict. We give thick descriptions of kinds of events that present as bot-bot reverts, helping distinguish conflict from non-conflict. We computationally classify these kinds of events through patterns in edit summaries. By interpreting found/trace data in the socio-technical contexts in which people give that data meaning, we gain more from quantitative measurements, drawing deeper understandings about the governance of algorithmic systems in Wikipedia. We have also released our data collection, processing, and analysis pipeline, to facilitate computational reproducibility of our findings and to help other researchers interested in conducting similar mixed-method scholarship in other platforms and contexts.

CCS Concepts: • **Human-centered computing** → **Computer supported cooperative work**; **Wikis**; **Empirical studies in collaborative and social computing**; • **Social and professional topics** → **Automation**; **Socio-technical systems**;

ACM Reference Format:
R. Stuart Geiger and Aaron Halfaker. 2017. Operationalizing Conflict and Cooperation between Automated Software Agents in Wikipedia: A Replication and Expansion of "Even Good Bots Fight". *Proc. ACM Hum.-Comput. Interact.* 1, 2, Article 49 (November 2017), 33 pages. https://doi.org/10.1145/3134684

**Code, data, and Jupyter notebooks available at:** https://github.com/halfak/are-the-bots-really-fighting
https://doi.org/10.6084/m9.figshare.5362216

## 1 INTRODUCTION

### 1.1 Introductory vignette: A bot-pocalypse?

*Between March 5th and 25th, 2013, one of the darkest periods in the robot history of Wikipedia occurred. An automated software agent called Addbot, operated by Adam Shoreland — an employee of the Berlin-based Wikimedia Deutschland foundation — committed the most aggressive bot-on-bot revert conflict event ever recorded. In a flurry of inefficiency and inefficacy, Addbot reverted 146,614 contributions other bots had made to English Wikipedia. It removed links between different language versions of Wikipedia*

> *articles, which had been automatically curated and updated by dozens of different bots for years. During a 20-day rampage, the bot annihilated their work and the work of their maintainers. The fact that such a massively destructive act could take place without being stopped is evidence that Wikipedia had failed to govern bot behaviors and that bots in Wikipedia are out of control.*

Or this is what we might conclude if we were to analyze activity from Wikipedia's database of edits without understanding the broader socio-technical context of what those traces mean in the Wikipedian community. If we operationalize "conflict" as one user account removing content previously added by a different user account, it is easy to find many cases between bots matching this criteria. This approach[1] was chosen by the authors of a paper published in *PLoS ONE* titled "Even Good Bots Fight: The Case of Wikipedia" [75] (or "EGBF"), which received major media coverage for reporting a massive number of cases where bots were reverting each other's edits, then concluding that Wikipedia had serious problems with how it regulates automated agents. However, if we look closely at Addbot's actions as they played out in the Wikipedian community, we instead see this case (and many more) as a productive, efficient automated operation performed by upstanding member of the Wikipedia community.

Addbot's removal of "interwiki language links" came after a massive effort coordinated through Wikidata[2], a meta-level database created to centralize language-independent data. For example, the links appearing on the sidebar of the "Japan" article on the English Wikipedia to "Japon" on the French Wikipedia and "日本" on the Japanese Wikipedia were once manually curated on the "Japan" page on the English Wikipedia. First human Wikipedians, then bots curated a long list of hidden links stored in the page text itself, which was parsed by the MediaWiki platform to display these links to readers. But with Wikidata, these links between all language versions were centrally curated, making the hidden links in each page's text redundant. Addbot's task was to remove these hidden links for pages that had their interwiki links migrated to Wikidata.

On Feb 14th, 2013, Adam Shoreland filed a proposal with the English Wikipedia's Bot Approvals Group to have Addbot take on this task, in line with English Wikipedia's Bots policy. It was the 32nd task he had applied for under this bot's account. According to Adam's request for approval, "The bot will use the wikidata api to determine [if] a EN article is already included on wikidata. If it is the bot will then find the list of links that are included along side it on wikidata and remove those from the article."[3] Within two days, various members of the Bot Approvals Group discussed the bot and then approved it for a full-scale run on English Wikipedia. Adam also made sure that the communities in other language versions of Wikipedia as well as Wikidata also approved of this task. To coordinate and bring visibility to the bot's work, he created a centralized wiki page for the link removal project, which was used to track approvals across various language versions of Wikipedia, as well as give information about when and where the bot would operate. Adam's application for the bot's operation was approved by many different wiki communities' Bot Approvals Groups, as well as given a global approval for wikis where the communities were too small to maintain a formal bot approval process on their own.

If we look at Addbot's actions out of context, it can appear to be the kind of out-of-control automated agent that indicates deep issues with Wikipedia's infamously open and decentralized governance model. Yet with the proper context, Shoreland's use of Addbot to remove redundant interwiki links is one of the best examples of a large-scale, well-coordinated, community-approved bot operation in an open, peer production platform. The bot's source code and operational status are clearly documented; it is operated in line with the various bot policies that different language versions can set for themselves; it is run by a known maintainer; and it is performing a massive task that would be a waste of human attention. The bot ran for a short period and completed the work as planned with no major objections or issue.

---

[1] *EGBF* analyzed data from 2001-10 and so does not include Addbot's work, but it would be cast as conflict by their method.
[2] https://wikidata.org
[3] https://en.wikipedia.org/wiki/Wikipedia:Bots/Requests_for_approval/Addbot_32

### 1.2  Even good bots fight? The broader implications of Wikipedian bot governance

This paper is a replication and expansion of "Even Good Bots Fight: The Case of Wikipedia" [75], which analyzed article edit metadata from several language versions of Wikipedia and claimed to find many cases of bots "fighting" with each other in Wikipedia. The EGBF paper operationalized bot-bot conflict through reverts, which is when one user account undoes another user account's edit. The paper (and especially the substantial mass media coverage it received) framed these cases of bot-bot reverts as deeply problematic automated agents out of control due to a lack of central governance and policy infrastructure. To summarize the EGBF paper using various quotes:

- "a system of simple bots may produce complex dynamics and unintended consequences"
- "conflict emerges even among benevolent bots that are designed to benefit their environment and not fight each other"
- "benevolent bots that are designed to collaborate may end up in continuous disagreement"
- "bots on Wikipedia behave and interact [...] unpredictably and [...] inefficiently"
- "disagreements likely arise from the bottom-up organization ... without a formal mechanism for coordination with other bot owners"
- "articles [...] contested by bots"
- "This is both inefficient as a waste of resources, and inefficacious, for it may lead to local impasse."

We are Wikipedia researchers who have been working on various issues around bots and automation for many years, and we were surprised and intrigued to hear these findings. While bot-bot conflict certainly takes place in Wikipedia for a variety of reasons, conflict at the size, scale, and severity the authors of the EGBF paper claimed has not been mentioned in almost a decade of multi-disciplinary scholarship on the governance of Wikipedia bots (and particularly the Bot Approvals Group) [13, 19, 24, 34, 43, 55, 65]. Previous literature has largely celebrated the successes of Wikipedia's approach to automation, which is based on the same principles as editing Wikipedia articles: decentralized consensus-building, scaffolded by formalized policies and processes. That said, several issues have been raised about the unintended impacts and effects of approved bots, particularly when human newcomers have their first contributions reverted by an anti-spam/vandalism bot. [27, 31]

If Wikipedia's existing bot regulatory system was leading to a substantial number of cases where approved bots were entering into prolonged conflict with each other over the content of Wikipedia articles, this would be a substantial cause for concern for Wikipedians. Furthermore, this issue has consequences outside of Wikipedia, as scholars, practitioners, and policymakers who are concerned about how to regulate algorithmic systems across application domains are looking to various real-world cases when thinking about what kinds of internal or governmental regulatory processes to implement. [2] Wikipedia is one of the oldest major user-generated content platforms (and has been working out various issues in using automation in user-generated content platforms since its inception in 2001). The successes and failures of Wikipedia's bot regulatory systems have broad implications. If Wikipedia's decentralized, consensus-based model was shown to lead to a substantial number of latent, unresolved conflicts with high inefficiencies, then this would discourage a similar model in other cases. So in this paper, we ask: *to what extent are bot-bot reverts* *in Wikipedia genuine conflicts where disagreements about how Wikipedia ought to be written were embedded in opposing bot codebases, versus cases like Shoreland's Addbot that reflect the opposite?*

## 2 LITERATURE REVIEW

### 2.1 Conflict in Computer-Supported Cooperative Work (CSCW) and social computing

Conflict is a complex issue, and CSCW scholars have extensively studied it in many areas [15, 56], including open source software, social media, and workplaces. As Kling [48] notes in an early foundational article on CSCW as a field, conflict is not merely inevitable, but can be generative:

> some group conflict is critical for identifying alternative lines of action and avoiding counter-productive conformity (groupthink), as long as conflicts are resolved constructively and with dignity. In practice, many working relationships can be multivalent with and mix elements of cooperation, conflict, conviviality, competition, collaboration, commitment, caution, control, coercion, coordination and combat. (p. 85)

More recent research on contemporary collaboration platforms has continued this multivalent approach, such as Filippova & Cho's interview-based study of open source software development communities [17]. They emphasize that conflict is multi-faceted and has many different sources and trajectories, including but going beyond debates over the codebase. The mere existence of conflict is not a cause for concern; we should focus on how conflict takes place, who it affects, and how it is resolved (or not). Literature in and around CSCW on conflict in distributed teams has identified various types of conflict, such as Hinds & Bailey's typology of task, process, and relationship conflict [38]. Task conflict is over what needs to be done, process conflict is over how tasks ought to be done, and relationship conflict is over interpersonal interaction styles and norms. These types of conflict can overlap, and one type of conflict can turn into another, but they are useful in distinguishing primary causes.

In this framework, if one bot developer believed that images without a properly-formatted fair-use rationale ought to be removed from articles, and other bot developer believed that such images ought to remain in articles, this would be task conflict. However, if both bot developers believed that such images ought to be removed from articles but had different ideas about what a properly-formatted fair use rationale looks like, this would be process conflict. And if a bot developer got angry at another bot developer and wrote a bot to undo all of their previous edits, this would be relationship conflict. These types of conflicts would all present as reverts, and it would be difficult, if not impossible to distinguish them using metadata alone. Yet they are strikingly different types of conflict that raise different issues.

### 2.2 Conflict in Wikipedia

Historically, Wikipedia researchers have often defined conflict through reverts [32, 33, 68, 81, 85], in which one user account removes another user account's edit to a page. This is typically defined as bringing it back to an exact previous state, called an *identity revert*. Other definitions include *partial reverts*, where part of an edit is removed, or *declarative reverts*, which is when a user leaves a note in an edit summary stating that they are reverting the edit (matching "rv" or "revert"). Kittur et al [45] reviews various approaches and notes that they are all "fairly consistent with each other" (p. 221) in terms of predicting editoral conflict between human editors — measured by Wikipedians' manual labeling of conflicts.

Kittur et al's original study [46] found that higher numbers of reverts were a strong predictor of conflict, but they also note that "reverts are complex actions that are an inherent part of the suggestion and negotiation that happens on discussion pages" (p. 7), citing longstanding scholarship on how conflict can be generative [21], see also [48]. They note a single revert is not indicative of conflict, but repeated back-and-forth reverts are indicative of an edit war and more negative outcomes. This is also how Wikipedians tend to define conflict. The "Bold, Revert, and Discuss" cycle (one of the most widely-cited essays [31] in English Wikipedia) states that editors ought to be bold in making changes, feel free to revert others' changes that they find unacceptable, then discuss the issue. Furthermore, not all reverts indicate conflict, and not all conflict takes the form of a revert. Wikipedia researchers have also studied how conflict takes place in talk page discussions, where editors can engage in hostile behavior without necessarily reverting each others' edits. [59]

## 2.3 Bot governance

As bots play a substantial role in Wikipedia, they have been extensively discussed in previous literature [13, 24, 34, 43, 55, 62, 65], even incidentally in the context of other topics [8, 19, 28, 51]. It is important to distinguish between: 1) *bot-bot conflict*, in which automated software agents get into "edit wars" with each other over the content of articles due to being programmed with opposing directives and 2) *conflict about bots*, in which human Wikipedians conflict with each other about what kinds of tasks ought to be automated in Wikipedia and how. Previous literature has not identified substantial amounts of the first kind of bot-bot conflict (which is what the EGBF paper claimed to find), although researchers have documented various discussions, debates, and controversies over the second kind of conflict over bots. Wikipedia's model of automation regulation is generally based on the same principles as editing Wikipedia articles: decentralized consensus-building, scaffolded by formalized policies and processes. Just like with editing articles, Wikipedians sometimes get into intense debates, conflicts, and controversies about whether the Bot Approvals Group ought to approve or deny a particular bot developer's application — and in some cases, about whether the BAG should rescind an approval for various reasons.

Geiger [24] discusses these processes of bot governance in Wikipedia, focusing on the history of Wikipedia's Bots policy and the operation of the English Wikipedia's Bot Approvals Group, which has existed in some form since 2004. Unapproved bots are not allowed to edit encyclopedia articles; they must get prior approval from the BAG for a specific, well-defined task. The BAG is a standing committee of bot developers and non-developers tasked with reviewing proposals about new bots in line with the community-authored Bots policy. Any Wikipedian editor is able to raise an issue about a bot to its operator, and if bot operators do not adequately respond to issues or concerns before their bot is approved, it will not be allowed to run. Operators of approved bots are also obligated to respond to issues and concerns raised to them, and failure to do so can get a bot's approval revoked by the BAG. The BAG serves as a clearinghouse for connecting bot developers and non-developers, although highly controversial cases can be adjudicated in larger dispute resolution processes, like the Arbitration Committee.[19] Furthermore, any Wikipedia administrator[4] has the ability to temporarily block any bot account from operating if they find it is malfunctioning or editing outside of its BAG approval scope.

In a case study of the German Wikipedia, Müller-Birn et al [62] symmetrically analyze bots and new features built directly into the MediaWiki platform, showing how they are both ways in which a developer "translates community consensus into practice" (p. 86). They discuss intense debates that are simultaneously about software and social norms. They also note a case in which a bot was approved for one task, but the developer expanded beyond the scope of their approval, prompting a controversy that had to be resolved. They also warn that "In both cases of algorithmic governance – software features and bots – making rules part of the infrastructure, to a certain extent, makes them harder to change and easier to enforce" (p. 87). Studies of newcomer socialization have also shown that some newcomers do not know if they are interacting with a human or a bot [18].

---

[4]Each language version approves its own administrators, who have various technical privileges and social expectations. English Wikipedia has about 1,250 admins, German has about 200, French as about 150, Chinese Mandarin has about 75.

To our knowledge, there has not been much qualitative or case study literature on non-English Wikipedian bot governance, with the exception of Müller-Birn et al's case study of German Wikipedia. Dozens of language versions of Wikipedia have their own bot policy and bot approval process or group (which are generally modeled on English Wikipedia), including Japanese, Italian, French, Romanian, Portuguese, Spanish, Japanese, and Mandarin Chinese. There is also a much stricter global bot approval process for the smaller language versions that do not have their own bot approval process (similarly modeled English Wikipedia), which Shoreland went through for Addbot in the opening vignette. Yet while there are still gaps in the literature on Wikipedian bot governance, we should note that the issue about to what extent bot-bot reverts constitute conflict was largely raised because the EGBF paper claimed to find what previous literature had not.

## 3 EPISTEMOLOGICAL OVERVIEW
### 3.1 How do we work with found data?

The study we present in this paper asks and answers our empirical research question about the extent to which bot-bot reverts in Wikipedia constitute bot-bot conflict. However, in this paper, we also frame our approach to answering this particular empirical question as an epistemological issue in computational social science: *when working with large-scale "found data" [36] of the traces users leave behind when interacting on a platform, how do we best operationalize culturally-specific concepts like conflict in a way that aligns with the particular context in which those traces were made?* In other words, how do we integrate the rich, thick descriptions of qualitative contextual inquiry in the practice of computational social science to arrive at a more holistic analysis? How can a research team both dive deep into the complexity of particular cases and scale up to the massive size of a dataset like those found in contemporary social computing platforms?

In the growing field of computational social science, researchers routinely begin with one or more large (but often thin) datasets generated by a software platform, which has recorded digital traces that users leave in interacting on that platform. Such researchers then seek to mine as much signal and significance from these found datasets as they can at scale in order to answer a research question. In more traditional qualitative social science methodologies that embrace induction and iteration, researchers continually seek to generate new data through interviews, observation (including participant-observation), focus groups, diary studies, surveys, archival research, or even experiments. Yet in contemporary computational social science, data mining rarely goes beyond the found dataset — a trend that has been critiqued by both scholars in the field of "critical data studies" [12, 40], as well as practicing computational social scientists.

For example, in a *Science* editorial on the shortcomings of the Google flu trends study, Lazer, Kennedy, King, and Vespignani — influential researchers and early champions of the promise of "big data" — critique this approach and "the often implicit assumption that big data are a substitute for, rather than a supplement to, traditional data collection and analysis" [54]. Furthermore, we do agree there can be some value in studying to what extent a convenient found dataset can, on its own, be used as a substitute for more traditional social science data. For example, it is interesting to ask whether social media posts predict elections [61] or whether Wikipedia edit activity predicts a film's box office revenue [60]. Yet when the goal is answering an empirical research question about a specific phenomenon a specific setting (particularly when making recommendations for practice and policy), we see no reason to limit ourselves to found data.

## 3.2 The perils and politics of decontextualization

In a related but distinct issue, scholars in the critical data studies literature have extensively discussed the problems of decontextualization, that is, when parts of social worlds are "datafied" [12, 78]. There is inevitably something left behind or incompletely captured in any transcription or translation of an entity or event in the world [10, 42, 70]. This is especially the case with databases from online platforms, which are primarily designed to keep the site running, rather than designed to be high-quality sources of rich social science data. As a result, these data tend to only give us a partial picture of complex social-cultural phenomena. For example, being a Facebook "friend" does generally represent some connection between two people, but it is dangerous to let it stand in for friendship, which is multivalent and differs dramatically across contexts. There is substantial literature and commentary on the pitfalls, dangers, challenges, and broader implications of data decontextualization (and other issues in using found data, particularly bias and ethics).

Scholars in and around CSCW have extensively studied and theorized how databases and platforms represent and reconstitute activity and users, from business information systems [66] to classification systems [6] to databases in general [1]. CSCW has a rich theoretical apparatus to discuss topics like the socio-technical constitution of identity in social media profiles [7], online dating profiles [3], and open source software repositories [57]. In a more direct critique, Tufecki [76] discusses issues in the interpretation of Twitter data, specifically critiquing the commonplace operationalization of popularity and influence for a given tweet using retweet data. She brings context from her fieldsites of various protest movements to show substantial differences in what people intend when they retweet a tweet. She also notes that there are many kinds of engagement on Twitter which are invisible in analyses that only use retweet data. We do similar kinds of work in our study of bot-bot reverts as conflict in Wikipedia.

## 3.3 Cooking data with care: veridical versus representational approaches

Our paper is aligned in spirit with these literatures, but we are not satisfied with simply critiquing the lack of context in large-scale computational analysis, then counterpoising more traditional qualitative methods of contextual inquiry. Instead, we present our own integrative, mixed-method approach to answering the EGBF paper's same general research question at the same scale about bot-bot conflict in Wikipedia. Our study is aligned with scholarship that brings together these ostensibly opposed approaches, particularly Neff et al's "critique and contribute" framework [64], which seeks "to support the integration of the perspectives of critical data studies into the organizational and social processes fundamental to doing data science."

We found Susan Leigh Star's classic text on *The Ethnography of Infrastructure* [73] to be especially helpful in distinguishing the different ways in which we can understand data collected by information infrastructures. Star discusses the "veridical" approach, in which "the information system is taken unproblematically as a mirror of actions in the world, and often tacitly, as a complete enough record of those actions" (p. 388). She contrasts this with seeing the data as "a trace or record of activities," in which the information infrastructure "sits (often uneasily) somewhere between research assistant to the investigator and found cultural artifact. The information must still be analyzed and placed in a larger framework of activities." (p. 388) Star cites Yates & Orlikowski's classic paper on the structuration of interoffice communication [86] as a good example of qualitative scholarship that relies heavily on found data, but not in the veridical approach. Yates & Orlikowski's interpretations of memos, e-mails, Lotus Notes documents, etc. are made within the local context in which they were created and circulated.

We take found data generated by software platforms like Wikipedia to be traces or records to be further interpreted, explored, linked to other records, and used as prompts in interviews. The benefits of taking a less veridical and more representational approach are clear for traditional qualitative methods, but we also found it useful in a quantitative and computational approach. For us, a revert is not universally mapped onto an instance of conflict, but instead serves as a starting point to explore what may or may not be conflict. We begin with the same database dumps of Wikipedia edit activity as the EGBF paper did, but we interrogate, refine, supplement, and extend this dataset in an iterative manner, integrating computational and ethnographic expertise to distinguish cases of conflict from non-conflict at various scales.

Our broader goal of this paper takes Geoff Bowker's quip about raw data to heart: "Raw data is both an oxymoron and a bad idea; to the contrary, data should be cooked with care." [5] At the broadest level, this paper is about how to cook and prepare a dataset with care. In our case, this means integrating ethnographic expertise about how bots operate in Wikipedia in the collection, cleaning, and analysis of bot-bot revert data. It involves going far beyond the original dataset to find contexts that have been left behind, including talking to people at the center of these cases, diving deep into discussion archives, and reconstructing how a particular pair of bots interacted with each other — and doing this with the background knowledge of how bots in Wikipedia operate based on years of ethnographic fieldwork.

### 3.4 Overview of the research study

Because of our broader epistemological goal, the way we present our research in this paper is less straightforward than a more standard computational social science paper. This is because the order in which we present our data, analyses, and cases is generally how we did our research — a rarity in computational social science.[5] While we could have presented this paper more linearly, it is important to stress that this reflexive process of exploration, iteration, refinement, expansion, validation, and synthesis is crucial to how we *cooked our data with care*. We also make heavy use of the second-person voice and reflect on our positionality (see section 6.1), rather than present a third person "view from nowhere" that de-emphasizes our interpretative role [35, 63].

We first manually explored some cases of bot-bot reverts identified by the EGBF authors in news media interviews and found a few cases that were certainly not conflict, like the one we presented in section 1.1. This prompted us to conduct a more thorough study, and we determined that the dataset of the study published by the EGBF authors did not retain enough of the edit metadata to ask the questions we had. We then generated our own dataset of bot-bot reverts from the published database dumps, using some improvements from standard approaches in the literature — improvements we learned from our time working in the Wikipedia community (section 4). We then conducted a high-level computational analysis using two metrics that are generally useful in distinguishing conflict from non-conflict: time to revert and reverts per pair of bots per page (section 5). These analyses showed that a substantial number of bot-bot reverts were undoing an edit made many months or even years prior, and that there were only a handful of cases where two bots reverted each other more than twice in a given article.

Once we had these two statistical distributions that indicated a substantial amount of non-conflict cases, we selected various cases for a deeper qualitative examination (section 6). To find these cases, we randomly selected cases from the extremes and middles of these distributions, as well as looked for specific cases we had encountered before in our previous experience working with, studying, and developing automated systems in Wikipedia. We soon found that bots approved for tasks working with interwiki links or redirect pages[6] kept appearing at much higher rates than other cases. We used a trace ethnographic approach to investigate what these bots were doing that presented as reverts and the extent to which they constitute bot-bot conflict. We depict these different cases in thick descriptions in sections 6.2 and 6.3. We also give detail about other kinds of cases of bot-bot reverts in the rest of section 6, which include some cases of bot-bot conflict.

---

[5] That said, our paper, code/data repository, and notebooks are still a partial narrative re-presentation of our research process. To present our full research process would likely be analogous to making a map the size of the territory; see [4].
[6] A page that forwards the user from one title (e.g. "Obama") to the canonical article (e.g. "Barack Obama").

As we had a set of identified types of bot-bot reverts that do and do not constitute bot-bot conflict, we wanted to scale back up to learn what proportion of our bot-bot revert dataset were made of these kinds of cases. We were not able to come up with any universal, context-independent, computational method of distinguishing between conflict and non-conflict using only the metadata of particular edits. Instead, as we present in section 7, we iteratively narrowed down our dataset by identifying types of bot-bot revert cases. Then we used pattern matching of edit summaries to classify these types in bulk, updating our dataset. With each pattern for matching a type, we manually investigated a random sample of cases to make sure it was only matching those types of cases. We continued to qualitatively investigate the unclassified cases: we would find a distinct kind of non-conflict activity or case of conflict, pattern match to classify the group in bulk, validate the pattern match by manually checking a random sample, then repeat the process with the unclassified cases until only a low proportion of the dataset was classified. Finally, once we had an augmented, labeled dataset, we could run additional statistical analyses at scale to investigate how these different types of bot-bot reverts compared (section 8).

## 4 GENERATING A DATASET OF ALL BOT-BOT REVERTS

In order to study the specific claims of the EGBF paper, we first needed a dataset of all bot-bot reverts, with the complete metadata for each edit. The dataset released by the EGBF authors only ranges from 2001-2010, does not contain full metadata for each edit, and was generated by software code that is not publicly available. To generate a new dataset, we needed to identify bot accounts and detect revert events. In this section, we discuss how we did so, what assumptions we made, and how our approach of dataset generation differs from the EGBF study.

### 4.1 Identifying bot accounts

Under almost all circumstances, Wikipedians use special accounts to operate bots, separate from the accounts that they edit under manually. This helps track the activities of a bot and differentiate it from a human actor. However, getting a list of bot accounts has long been a challenge for Wikipedia researchers. Our strategy differs from the EGBF paper, which had a novel approach of searching for usernames containing "bot" plus using a status page of bots on English Wikipedia, followed by manual checking of user accounts. The EGBF approach may miss bot usernames not on the status page (which is no longer actively maintained) that do not include the English word "bot" in their usernames — like "ArticleGrinder" and "苏筱弯." Their dataset of bot names is also not public.

Our strategy also differs from the standard way Wikipedia researchers have identified bots, which is through "user groups" (also called the "bot flag") [23, 44, 60, 67, 83]. We use not only the bot user group, but also a new automated approach based on the user_former_groups table and the "Bot user" cross-wiki category. Using our strategy, we find 6,522 current and former bot accounts across all language versions, and the generation and publication of this dataset of bots across language versions is an independent contribution of this study to the Wikipedia research literature. The dataset of bot accounts and the code to automatically update this dataset is available on our GitHub repository[7] and on Figshare [26].

The bot user group has limitations that have not been discussed in previous literature. MediaWiki, the software used by Wikipedia, manages user rights and restrictions using "user groups," also called flags. For example, Administrators have more rights than regular users and are assigned the "sysop" group/flag. Bots have rights that an average user doesn't have and are assigned the "bot" group/flag. MediaWiki tracks user groups using the user_groups table in its database, which provides a convenient mechanism for obtaining a list of currently active bots. However, bot accounts are often removed from the bot user group (or "unflagged") when they are no longer active. There was no history of that removal until November 2011, when MediaWiki 1.18 implemented the user_former_groups table that logged all future changes to user groups. We queried and merged bot accounts from both the user_groups and user_former_groups database tables. Through querying English, German, Portuguese, Spanish, Chinese, Japanese, and French Wikipedias, we identify 2,529 unique bot accounts.

---

[7] https://github.com/halfak/are-the-bots-really-fighting

However, there are many bots flagged before 2011, as well as bots that were never flagged at all (some Bot Approvals Groups will specifically approve a bot but decide to not flag it, because flagging makes bot edits not appear in recent changes feeds by default). To capture these bots, we relied on categories of bot accounts curated by editors on each language version of Wikipedia. Wikipedians on the English Wikipedia curate a category titled "Category:All_Wikipedia_bots" containing active and inactive bots on en.wikipedia.org, while Wikipedians on the French Wikipedia curate a similarly-functioning category for fr.wikipedia.org titled "Catégorie:Bot Wikipédia." 154 language versions of Wikipedia (as well as commons and meta.wikimedia.org) have such categories, which are each linked to the Wikidata item "Q3681760." By querying first Wikidata for all of these language-specific categories, then all of the pages appearing within these categories on each wiki, we identify an additional 3,993 bot accounts.

### 4.2 Identifying reverts

In order to identify reverts in Wikipedia, we made use the of 'mwreverts' python library that is designed to extract revert events from several types of MediaWiki datasources [30]. For our analysis, we used the "stub-meta-history" XML database dumps provided by the Wikimedia Foundation.[8] For German, Portuguese, Spanish, Chinese, Japanese, French, and English Wikipedias, we used the April 20th, 2017 data dumps (archived at [20]), then truncated to December 31st, 2016 for consistent year-by-year statistics. We used the "mwreverts dump2revert" utility to identify identity reverts by SHA1 hashing, then filtered the reverts so that only the closest reverted revision to the reverting revision was included (as described in the EGBF paper). Finally, we filtered the dataset for reverting and reverted usernames that match any of the names from our unified set of 6,522 bot accounts.

### 4.3 Underlying software used

This analysis was conducted in Python [79] and R [71], using Pandas dataframes [58] and data.table [14] for data parsing and transformation, SciPy [41] and NumPy [77] for quantitative computations, and Matplotlib [39], Seaborn [82], and ggplot2 [84] for visualization. Analysis was conducted in Jupyter Notebooks [49] using the IPython [69] and IRkernel kernels. A full specification of our computational environment (including version numbers) can be found in our GitHub repository.

## 5 EXPLORATORY ANALYSIS: ARE THERE SIGNALS OF BOT-BOT CONFLICT?

Once we had our dataset, we first wanted to run some descriptive statistics and determine if the rate of bot-bot reverts had been changing over time (section 5.1). We also wanted to see if there were signals of bot-bot conflict across the entire set of bot-bot reverts in Wikipedia. To do this, we needed metrics that would both be meaningful in aggregate distributions, as well as help us identify cases to explore in detail. We used two metrics: time to revert (section 5.2) and reverts per pair of bots per page (section 5.3). Both metrics provide strong evidence that a large proportion of bot-bot reverts were likely not conflict, but something else to be further investigated. This analysis helped us identify various issues and further avenues for investigation. We return to these metrics later in section 8 when we present an analysis of bot-bot reverts classified by type based on edit summary comments.

---

[8] https://dumps.wikimedia.org

### 5.1 How many bot-bot reverts are there and how has their rate changed over time?

We collected a total of 924,945 bot-bot reverts from January 2001 to December 2016 across all pages of seven language versions of Wikipedia. Our base dataset includes reverts from pages across all "namespaces" of Wikipedia: articles are in the main namespace or namespace 0, article talk pages are prefixed with "Talk:" and in namespace 1, User pages are prefixed with "User:" and in namespace 2, and so on. Like the EGBF paper, much of our analysis is filtered to articles only, even though about 39% of bot-bot reverts take place outside of the main/article namespace (we discuss bot-bot reverts in articles versus other namespaces in section 6.4.4).

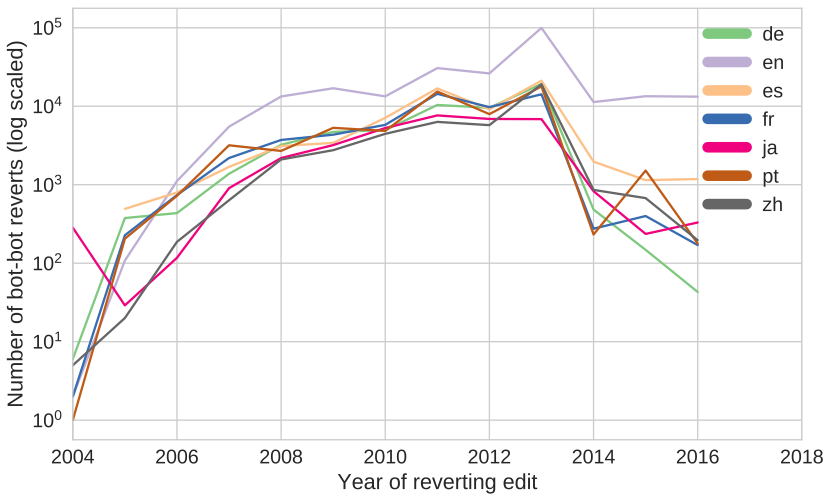

Fig. 1. The number of bot-bot reverts per language per year, for articles only

Figure 1 is a log-scaled plot of the total number of bot-bot reverts per language by year (for only articles). This shows that bot-bot reverts grew and peaked in 2013 across all languages, then have declined since. In English Wikipedia articles, bot-bot reverts grew from 13,512 in 2010 to 99,694 in 2013, but have declined to 13,245 in 2016. Other language versions have dropped substantially, such as Mandarin Chinese Wikipedia's peak of 19,089 reverts in 2013 to 675 reverts in 2015 and 196 in 2016. We also used our dataset of bot accounts to collect the total number of edits bots made across each language version, then found the proportion of bot edits to articles that were bot-bot reverts. This was between 0.5% and 1% for all languages in our dataset.

### 5.2 Time to Revert

*5.2.1 Introduction.* One of the simplest factors to examine when trying to separate conflict from non-conflict is *time to revert*, or the time between when a reverted and reverting edit is made. Time to revert is a longstanding metric used in Wikipedia research, including conflict [80] and automation. [25, 31] If one bot makes an edit, then another bot reverted that edit five minutes later, this would be far more likely to be conflict than if the revert was made five years later — all other factors equal. One of the foundational principles of Wikipedia is that consensus can change: an edit can be made that conforms to policy, then reverted a year later because consensus changed. In fact, bots are frequently requested and built to update changes in project-wide policy, including the case of Addbot referenced at the beginning of this paper. As such, *time to revert* does not exclusively distinguish conflict from non-conflict, but high time to revert values indicate a different kind of activity that is less likely to be the kind of conflict of concern in Wikipedia.

*5.2.2 Time to revert analysis.* When we plotted the distribution of times to revert for articles across languages in a Kernel Density Estimation[9] (Figure 2), we found a substantial amount of bot-bot reverts were undoing an edit the other bot had made months, even years after. Different language versions varied, but for articles across all languages in our dataset for all years, the median time to revert was 62 days. English had the highest mean and median (mean=355 days, median=138 days), Japanese had the smallest (mean=107 days, median=13 days).

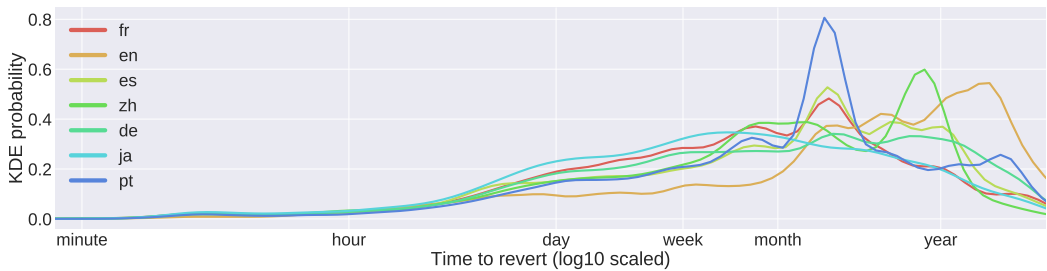

Fig. 2. Kernel density estimation showing the distribution of time to revert values for bot-bot reverts per language, articles only. Y axis is the probability a bot's edit was reverted by another bot after X time.

When we broke out the distribution of times to revert by both language and the year the reverting bot reverted the original bot edit (Figure 3), we found that times to revert have generally been rising substantially. For example, in the English Wikipedia, median time to revert in 2008 was 26 days, but had risen to 299 days by 2016. This possibly results from two factors: first, the removal of interwiki language links made redundant by Wikidata post-2012, many of which were added by other bots years ago; and second, the decreased incidents of bot-bot conflict as various language versions gained stronger and more established bot policies and bot approval venues.

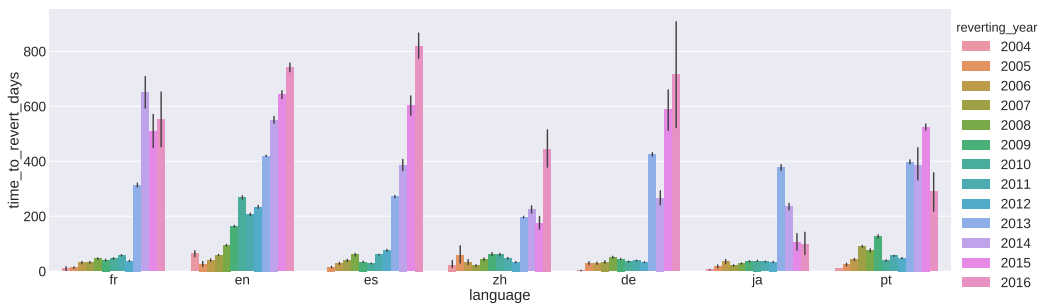

Fig. 3. Mean time to revert in days for all bot-bot reverts by language by year, articles only. Error bars are 95% confidence intervals.

---

[9]A Kernel Density Estimation is a non-parametric way of visualizing a distribution; effectively a smoothed histogram.

## 5.3 Reverts per page

In reading the initial EGBF paper and mass media coverage, many representations of the bot fights imagined two bots having lots of interactions on a single page — bot A makes a change, bot B reverts it, bot A reverts bot B's revert, and so on — each triggering each other's action ad infinitum. If two bots are having many revert interactions on a single page, this would be strongly indicative of some type of conflict. In order to explore this potential bot dynamic, we computed the counts of interactions between unique bot pairs (the reverting and reverted bot's username) on each page. This method was extremely useful in identifying various cases that, upon qualitative investigation, were the most intense "bot fights" or edit wars in which two bots were contesting the content of articles. Similarly, bot-bot reverts that had only one bot-pair interaction per page were generally the kinds of routine, non-conflict work like removing obsolete interwiki links. We also measure the duration (time between first and last bot-bot revert) of a bot-pair revert event in an article to get a sense for how long these conflict-like interactions play out before being resolved.

Figure 4 is a log-scaled histogram plotting the number of articles in English Wikipedia where $n$ reverts are made by the same pair of bots on the same article. Out of the 244,793 bot-bot reverts to articles in English Wikipedia, 228,198 bot-bot reverts per pair of bots only happen once in an article. This means that 93.2% of bot-bot reverts were not reciprocated or otherwise responded to by the bot that was reverted. 13,012 bot-bot reverts (5.3% of all bot-bot reverts) happen twice in a single article, meaning there are 6,506 cases where a bot reverted another bot, then the reverted bot reciprocated with a revert of their own on the same article. Only 3,583 bot-bot reverts (1.46% of all bot-bot reverts) were part of a case where two bots reverted each other more than twice in a single article. The few cases with high values on the X axis in figure 4 indicate an extended edit war between bots on the same article: our max was two cases where the same two bots each reverted the other 41 times on the same article. We discuss this case in section 6.5.3. We also found a similar pattern across all language versions: in total, out of 561,936 bot-bot reverts to articles, 528,104 bot-bot reverts (93.9%) were not reciprocated, 25,528 (4.5%) were part of a case where the reverted bot reciprocated with a revert on the same article only once, and 8,304 (1.48%) involved more than two reverts per pair of bots on a single article.

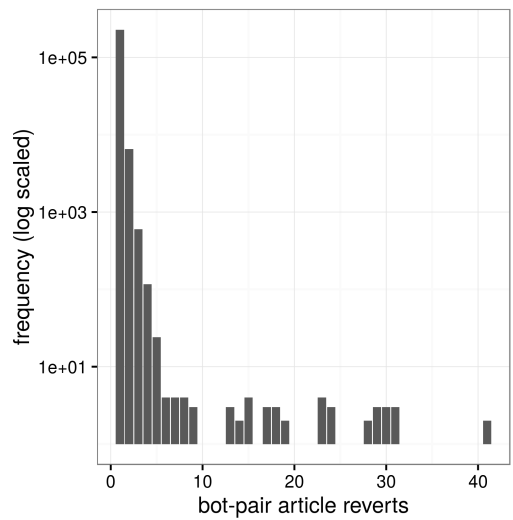

Fig. 4. A histogram of the frequency of bot-pair interactions for each article in English Wikipedia, scaled logarithmically.

## 6 QUALITATIVE DESCRIPTIONS OF VARIOUS TYPES OF BOT-BOT REVERTS

At this point in our investigation, based on the quantitative findings presented in section 5, we strongly suspected that the overwhelming majority of bot-bot reverts did not constitute conflict. Most bot-bot reverts took place months or even years after the original edit, and most were not reciprocated by the reverted bot — however, a small number did exhibit one or both of these indicators of conflict. So our next steps were to ask: *What kinds of work are these bots were doing?*

*What were the bot's developers intending the bots to do? And to what extent do these bot-bot reverts constitute conflict within Wikipedia's policy environment?*

## 6.1 Principles of trace ethnography and thick descriptions of databases

These questions cannot be answered with metadata alone, as they require a deeper situated understanding of how work takes place within Wikipedia, particularly in the context of how work is delegated to bots and other automated software agents. As such, our approach drew on the principles of trace ethnography [29], which is based on a researcher learning how to follow and interpret transaction log data as part of the lived and learned experience of a community. Trace ethnography is not "lurker ethnography" done by someone who never interviews or participates in a community. It is instead based on the kinds of interpretive expertise (or "trace literacy" [18]) a member of a mediated community gains as they learn how to participate in that community. As Geiger & Ribes discuss in their study of "vandal fighters" [28], Wikipedians have developed a rich repertoire of interpretive practices, which they routinely rely on when trying to figure out what has taken place on the site. In many ways, our explications of what happened with the various bot-bot reverts in this section are similar to the fact-finding section of a decision by the Arbitration Committee — a quasi-judicial body that acts as a kind of Supreme Court for the English-language Wikipedia, adjudicating complex and high-level disputes, including several cases around bots.

Both of us each have over 10 years of experience participating in the Wikipedian community as volunteer contributors, which includes mediated participation on wikis, mailing lists, and chat rooms, as well as attending local, regional, and global Wikipedia/Wikimedia meetups and conferences. Our participation in this community includes writing and editing encyclopedia articles, working as a "vandal fighter" to identify malicious contributors, discussing and debating a wide range of micro-to-macro level issues, and helping develop various automated tools and bots. We also have respective expertise as a qualitative ethnographer and a computational social scientist who have extensively studied the impacts and effects of various tools and bots in Wikipedia. We have also both been employed by the Wikimedia Foundation in various research roles: both of us worked as WMF research interns in 2011; one of us worked as a part-time research consultant from 2011-2012; the other has worked as a full-time research scientist from 2013 to the present. Most of our experience is on English-language wikis and venues, as English is our primary language.

Based on our expertise, we are able to start with the revision ID of an identified bot-bot revert in our dataset and follow other traces to find other contextual information, such as: what was changed in the revert, the bot developer's summary of what the bot was doing at the time, the other edits that both bots had made, both bots' Requests for Approval before the Bot Approvals Group (if any), the various talk pages where Wikipedians would raise and resolve the conflict (in cases of genuine conflict), and the various policies and guidelines in force at the time of the revert. Like other veteran Wikipedian contributors, we also know many of the idiosyncrasies about how the MediaWiki platform and the underlying database operates, like how page moves and renames are handled in revision histories; we know what does not get captured in logs and revision histories; and we know how the platform and database has changed over time. Finally, we independently checked each other's interpretations and assumptions across cases, then in some cases also sought out Wikipedian volunteers to triple-check that our interpretations were accurate.

We begin our first description of a case by presenting a "thick description" [22] of an extremely common issue in Wikipedia that presents as bot-bot reverts: cleaning up links between pages. These descriptions are long, but just as Geertz sought to provide rich contextual detail that distinguishes a wink from a blink from a twitch, we seek to provide detail that helps distinguish conflict from non-conflict. The extended length of the sections on link cleanup is also justified because these cases around fixing double redirects and managing interwiki links make up 95.8 % of the bot-bot reverts to articles across our dataset, as we show in section 7.

## 6.2 Fixing double redirects in Wikipedia

### 6.2.1 The double redirect problem.

One case of bot-bot reverts that was mentioned in several mass media articles about the paper (but was not as fully detailed in the paper itself, which did not provide examples of content allegedly "contested" by bots) was the ostensibly antagonistic interactions between two bots named Xqbot and DarknessBot. These two bots fix double redirects, which arises in Wikipedia around the titles of Wikipedia articles. Wikipedians have disagreements over what the canonical title for a given topic ought to be. Should the main article on Japan-U.S. relations be titled "Japan-U.S. relations" or "U.S.-Japan relations?" Or should the main article on the 44th president of the U.S. be titled "Barack Obama," "Barack H. Obama," or "Barack Hussein Obama"? Wikipedians must resolve these issues on a daily basis, across the millions of different articles across hundreds of language versions. As the example of Obama's article's title indicates, there can be extensive conflict between human Wikipedian editors over the canonical title of an article. As Hill and Shaw [37] discuss in their paper on redirects, they are an understudied but foundational aspect of Wikipedia, "play[ing] an important role in shaping activity" (p. 1).

The double-redirect problem also arises because of decade-old decisions made about Wikipedia's underlying software platform and database. To deal with alternative spellings, formatting, or expressions of article titles, early wiki platform developers created redirects, which are pages that anyone can edit (just like articles) that silently redirect readers to another page. Every Wikipedia article on a topic has a single canonical title, which is the basis for the article's URL. Then there are dozens or potentially hundreds of other pages that contain no text other than some wiki-code that redirects the reader to the page where the canonical article is located. So the article about Barack Obama is titled "Barack Obama" and is accessible at enwp.org/Barack_Obama, but there are many other pages that redirect to this, like enwp.org/Barack_H._Obama and enwp.org/Obama. Most major language versions of Wikipedia have a Manual of Style with a section on article titles guiding decision-making about these issues, as well as a specialized process for discussing redirects and canonical names, called "Redirects for Discussion" in English Wikipedia.

In Wikipedia, redirects are a continual site of both genuine disagreements and malicious vandalism. In line with Wikipedia's "anyone can edit" principles, anyone with a Wikipedia account that is "autoconfirmed" (i.e. has been registered for more than 4 days and made at least 10 edits without getting banned) can unilaterally rename any non-protected page. Only admins can change the enwp.org/Obama redirects, because it has been protected after substantial edit wars over the name. But any autoconfirmed editor can unilaterally can rename the titles of the overwhelming majority of Wikipedia articles. Any other autoconfirmed editor can unilaterally undo their rename, which is how Wikipedia's editorial model generally works. If two or more editors get into an edit war, then they (or other non-involved parties) can request dispute resolution and make a formal proposal at Redirects for Discussion. This is a normal part of how Wikipedia has operated for over a decade.

One technical problem arises when a human Wikipedian decides to rename and move a page that had other redirect pages pointing to it. Those redirects will still be pointing to the now-old title, which is called a double redirect. Double redirects are inefficient and mess up the redirect link graph (e.g. what redirects to what), which causes problems both for readers and for people who use this data for research applications. Many bots have been created and approved to solve this problem by automatically finding and fixing double redirects, which is a relatively easy computational task. It is such a common task that code for fixing double redirects has been built into the most popular open source Wikipedia bot framework, pywikibot, so that a bot developer can easily turn any bot into one that also fixes double redirects as it goes about its main task.[10] In fact, many of the bots approved to do double redirect fixing use this exact same code, even though they are operated independently by different developers. Yet as we show in the next section, the work of fixing double redirects sometimes demands that bots revert each other's edits. This does not indicate conflict (either intentional or unintentional) between the bots and/or their developers; instead, the bots are collaboratively cleaning up after humans who have reverted each other over the canonical title of an article.

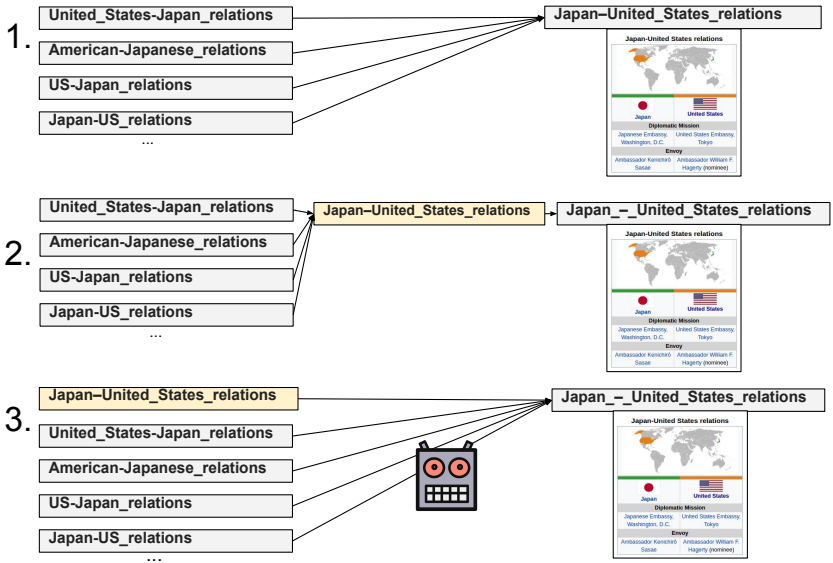

Fig. 5. A diagram illustrating the redirect graph of the case in this section: (1) the initial state of the redirect graph, (2) a double redirect being created by The Transhumanist's renaming, then (3) a bot fixing the double redirect

*6.2.2 Case: Xqbot vs. DarknessBot on the Japan-United States relations article.*
In this section, we give a detailed account of one such case between two bots that were mentioned by the EGBF authors in media interviews as an example of bot-bot conflict. This case examines the revision history for the page titled "Japan-United States relations," which is accessible at enwp.org/Japan-United_States_relations. The relatively short revision history (the relevant section in Table 1) shows what appears to be bots reverting other bots. Xqbot makes an edit, DarknessBot reverts that edit taking it back to the original state, then Xqbot reverts DarknessBot, putting the page back to the version it was in before. The page at "Japan-United States relations" is not the main article on this topic, but rather a redirect to the main article, which is currently titled "Japan–United States relations" — using a longer en-dash for the dash.

At present, there are 45 articles on the English Wikipedia redirecting to the main article on "Japan— United States relations," so that if you go to enwp.org/US-Japan_relations or enwp.org/Relations_between_USA_and_Japan or others, you will be redirected to enwp.org/Japan–United_States_relations. This article was originally created as "Japan-United States relations", with a shorter hyphen (or minus sign) for the dash. This article existed in with this same title for a few years, until en-dashes became popular. The Manual of Style was updated, and shorter hyphens began to be replaced with the longer en-dashes all across the English-language Wikipedia.

---

[10] https://www.mediawiki.org/wiki/Manual:Pywikibot/redirect.py

Table 1. Partial timeline of events on the page at enwp.org/Japan-United_States_relations

| Date of edit | Last edit by | Contents of page |
| --- | --- | --- |
| 2008-07-06 | The Transhumanist | #REDIRECT [[Japan–United States relations]] |
| 2009-02-15 | Xqbot | #REDIRECT [[Japan – United States relations]] |
| 2009-11-15 | DarknessBot | #REDIRECT [[Japan–United States relations]] |

On July 6th 2008, a human Wikipedian editor named The Transhumanist renamed the article title from "Japan-United States relations" to "Japan–United States relations", in line with Wikipedia's style guide at the time. The main article was now to be found at enwp.org/Japan–United_States_relations. The original main article was renamed in the database's page table, which has one row per page. It kept the same numeric page_id (and therefore the same revision history), but changed the page_title, which is what appears to anyone browsing in the web interface. A new page was created in the database with a new page_id, and the title was set as "Japan-United States relations." The content of this new page was a single line, containing the special wikimarkup text:

#REDIRECT [[Japan—United States relations]][11]

This made the page at enwp.org/Japan-United_States_relations into a redirect, automatically sending readers to the version with the longer en-dash if they navigated there. The next year, on February 15th 2009, a different human Wikipedian editor renamed and moved the page again, adding a space between the hyphen, so that it became "Japan – United States relations" and was accessible at enwp.org/Japan_–_United_States_relations.[12] The editor quoted the relevant part of the style guide at the time, which stated that "All disjunctive en dashes are unspaced, except when there is a space within either one or both of the items." This is a relatively low-conflict action, but like all title renames, it caused a technical problem: the original page (with a minus sign for a dash) is still redirecting to what was once the main article, which uses an en-dash but no spacing. This is called a double redirect, as someone who goes to enwp.org/Japan-United_States_relations will be redirected twice: first to enwp.org/Japan–United_States_relations, which then redirects them to the new version at enwp.org/Japan_–_United_States_relations

Double redirects are generally considered undesirable for many reasons. However, this is easy to computationally detect and fix: the bots find one of these double redirects with the pattern $A \rightarrow B \rightarrow C$ and make a redirect from $A \rightarrow C$. There can be many different bots that all independently perform this same task operating at the same time in Wikipedia, because it does not matter which one gets to it first. Less than five hours later, Xqbot detected an double redirect issue and got to work. It edited the page at enwp.org/Japan–United_States_relations to redirect directly to the new version of the page (with spaces between the en-dash) instead of the older one.

Almost exactly seven months later, on 15 September 2009, a third human Wikipedian editor renamed and moved the main article again, going back to the version with no spaces between the en-dash. The editor referenced the Manual of Style, implying that the policy had changed. [13] We now have another double redirect: from the page at enwp.org/Japan-United_States_relations to the page at enwp.org/Japan_–_United_States_relations, which redirects to the main article at enwp.org/Japan–United_States_relations. This was detected by another bot, DarknessBot, which is programmed to also find and fix double redirects. Using the exact same code (from pywikibot) as

---

[11] https://en.wikipedia.org/w/index.php?title=Japan-United_States_relations&action=edit&oldid=224031785
[12] http://tinyurl.com/wikimovelog1
[13] http://tinyurl.com/wikimovelog2

Xqbot, based on the same principle that double redirects can and should be fixed automatically ASAP, DarknessBot goes into action. It edits the enwp.org/Japan-United_States_relations article and makes it redirect to enwp.org/Japan–United_States_relations.

The diff of DarknessBot's 15 Sept 2009 edit[14] is a bot-bot revert by all standard computational measures that are commonly used in the analysis of Wikipedia log data (as are the 24 other reverts that took place over a few minutes to handle all the relevant redirect articles). Xqbot edited an article to so it was in a certain state, and Darknessbot took the article back to the identical state it was in immediately before Xqbot made its first edit. However, given the context, this kind of event should not be taken as a sign of bot-bot conflict, much less conflict between Wikipedians about bots. The only thing that can be framed as conflict is between two human Wikipedian editors who had different ideas about what the canonical name of the article should be. The bots should instead be seen as collaborating to keep the redirect graph clean, in the wake of human editors who disagree about page titles. Yet even the humans in this story should not necessarily be seen as in conflict, given the length of time that elapsed and how Wikipedia's own Manual of Style can change over time as Wikipedians change their positions on a multitude of minor details.

### 6.3 Migrating interwiki links

Another major type of bot work that appears as bot-bot reverts are a variety of work involving interwiki links, which are the links between language versions of Wikipedia on the same topic. The opening vignette featuring Shoreland's Addbot was about a bot that removed interwiki links after new integrations with Wikidata made them unnecessary for most purposes. Before migration to Wikidata took place, bots frequently did work cleaning up interwiki links, which is similar to how the double redirect bots clean up double redirects. The problems are of the same type, as they arise because human Wikipedians rename pages back and forth. However, they have another layer of complexity because bots are curating links across language versions. For example, when the article titled "Japan–U.S. relations" on English Wikipedia was renamed to "Japan – U.S. relations", bots had to update all the interwiki links on all the other language versions to point to the new canonical title. When the title was renamed back, bots had to update these interwiki language links again, which present a bot-bot revert across all language versions that had active interwiki updating bots operating. These should not be considered conflict for the same reasons as double redirect fixing: they stem from human editors making decisions about how to rename pages, with the bots doing precisely the work they were authorized and delegated to do in order to keep Wikipedia working behind the scenes.

### 6.4 Other cases of non bot-bot conflict reverts

#### 6.4.1 Maintenance, orphan, and protection templates.

Another common case that presents as bot-bot reverts is the updating of article notification templates. These templates are rendered at the top of an article to tell readers and editors that a page is protected from editing or that it has some issue. When an administrator temporarily locks a page from public editing, there are bots that will add a page protection template automatically. When the period is over, there are other bots that will remove this now out-of-date notice. This presents as a revert — and for pages that are frequently protected and unprotected, it can present as a long-term revert war — but it is a prime case of bot-bot collaboration, rather than bot conflict.

Other cases of bot-bot reverts over page templates include orphan templates, which several bots add when they detect that a page has no incoming links from other wiki pages. There are other bots that will remove an orphan template if they detect that a page now has incoming links.

---

[14] http://tinyurl.com/wikidiff2

This apparent revert cycle can also present as a revert war, if for example, an article only has one incoming link, but that link is removed and replaced, or the linking article is deleted and recreated. Like page protection templates, this is a prime case of bot-bot collaboration, not conflict.

Other "maintenance" templates exist in Wikipedia for informing readers that specific types of work need to be done on the article. If an article is not placed in any categories, then there are bots that will leave banners asking editors to categorize it. There are also bots that will remove the categorization template banner if it is further categorized. Alternatively, if an article has too many {{citation needed}} statements, there are bots that will add a banner template stating that the article needs reference work — and there are bots that will remove this template if the {{citation needed}} references are removed. Even though this kind of back-and-forth activity is the perfect definition of a revert, it is also a prime case of bot-bot collaboration. To see these kinds of reverts as conflicts ignores the context in which these templates are added and removed from articles.

### 6.4.2 Moving categories.

Page categorization in Wikipedia is complex and labor-intensive, providing another site for bot-bot reverts that do not generally indicate conflict. A page's current categories must be stored in the page text (like interwiki links were before Wikidata). To add an article to a category, an editor adds "[[Category:American women novelists]]" to the page source, which is then parsed by the MediaWiki platform to update various tables in the database. Categorization can also be contentious, with Wikipedians frequently disagreeing over the proper categorization schema as well as the names of categories (see the American women novelists controversy [16]). Like redirects, Wikipedians frequently debate categorization, with an established policy and process in the English Wikipedia for resolving these issues.

There have been several bots created to support categorization work. In the English Wikipedia, a set of bots are programmed to look for traces that administrators can leave after a Redirects for Discussion case is resolved with a decision to rename a category. These bots will then implement the decision by updating the category links in the article text. If a page is renamed back and forth, with the decision implemented by different bots, then this can present as a bot-bot revert. Cydebot was the first bot to implement this code, which the developer open sourced and built into the pywikibot framework for other bots to implement. Several additional bots have been approved to also carry out this work and run independently. Like with fixing double redirects, it does not matter which bot gets to do the category move first.

One case we found around moving categories illustrates the socio-technical complexity of bot-human interactions in Wikipedia. An approved bot named RussBot is programmed to search for categories that have a redirect tag on their page description, which is a somewhat common practice. For example, before 20 Aug 2009, articles about Super NES games were to be categorized in [[Category:Super NES games]], with [[Category:Super Nintendo Entertainment System games]] redirecting to the shorter version. If an editor categorized a page in the longer title by mistake, RussBot would re-categorize it in the correct title.

A discussion took place on 20 Aug 2009 in which Wikipedians decided to use the longer name as the default category.[15] An administrator placed a particular template on the Redirects for Discussion page instructing Cydebot to move pages from the shorter to the longer category. However, the administrator forgot to update the page for [[Category:Super Nintendo Entertainment System games]] as they should have, which was still redirecting to [[Category:Super NES games]]. Because of this, RussBot re-categorized all the pages back again, which Cydebot soon automatically reversed. The developer of RussBot discovered this and removed the category redirect on their own. The developer then left a message to the developer of Cydebot, raising the issue and recommending that Cydebot be updated to make sure this kind of event does not happen in the future. These two bots are programmed to collaborate on redirect tasks based on the instructions from a human Wikipedian, but because the administrator updated Cydebot bot's directive but not RussBot's, both ended up in a short revert war with each other. Thankfully, this issue was quickly identified and fixed due to RussBot's developer keeping a close watch on what their bot was doing.

---

[15] http://tinyurl.com/cydebottalk

*6.4.3 Bots editing with "per" justifications.*
Another common bot type are single purpose bots that are written to implement a decision reached by Wikipedians that applies to many articles in a category or even across the encyclopedia. For example, Wikipedians often decide to delete an image (usually for copyright reasons) and a bot will automatically remove the wikimarkup in articles that inserts the image in the page. Deleted categories are also removed from articles in bulk using bots. When these events occur, the common convention in English Wikipedia is to leave a comment in the edit summary stating what the bot is doing "per" a particular venue or discussion page. We found similar patterns in other language versions, such as "según" and "conforme pedido" in Spanish and "suite à discussion" in French. These cases can present as bot-bot reverts because bots may have been approved to automate some change to an article by a consensus at the time, which can change. In such situations, a new discussion and decision is made, then a bot is written to implement it, which undoes the work of the previous bot. This should not be considered bot-bot conflict, although it can indicate conflict between Wikipedians about what bots ought to do.

*6.4.4 Discussions and queues outside of the article namespace.*
While our paper is generally limited to encyclopedia articles (pages in namespace 0; as did the EGBF paper), we analyzed all pages across 7 language versions of Wikipedia. These other pages include discussion pages for individual articles, meta-level discussion spaces, administrative processes, sandboxes, files, and category descriptions. Bot-bot reverts are prevalent in talk pages, and while we did not focus extensively on them, we found many of the same kinds of patterns. Bot-bot reverts were prevalent on process pages like English Wikipedia's "Administrator Intervention against Vandalism" (AIV), where humans and bots leave requests to admins to investigate alleged vandals. Counter-vandalism bots like ClueBot NG will leave requests upon reverting egregious vandalism, and the HBC AIV Helperbots remove requests from this page once an admin has made a decision. We found 54,946 bot-bot reverts on this page alone, which is also an excellent case of bots collaborating to help humans get work done. Other pages outside of the article namespace with high bot-bot revert counts include bot-curated administrative request queues like "Usernames for administrator attention" (18,918 bot-bot reverts) and "Suspected copyright violations" (3,659 bot-bot reverts). Like template work, these cases present as bot-bot reverts, but they are part of an established human-bot workflow. We have also learned of conflicts over bots that may also include bot-bot conflicts on pages for files and categories, but we must leave such cases for future research.

## 6.5 Cases of bot-bot conflict over article content

*6.5.1 Mathbot and Frescobot over link syntax.*
One case we are confident to classify as bot-bot conflict is between Mathbot and FrescoBot. FrescoBot is programmed to fix link syntax mistakes. In wikimarkup, a link to Xenocrates is made with [[Xenocrates]], but the displayed link can be changed with the syntax [[Xenocrates | see the Xenocrates article]]. FrescoBot was programmed to find cases where the link and the displayed link were identical, such as [[Xenocrates | Xenocrates]]. Mathbot was a bot that curated 54 alphabetical index pages for Lists of Mathematicians and List of Mathematics Topics. If an article was added to one of several mathematics categories, MathBot would add it to the appropriate list. However, it would add links in the syntax of [[Xenocrates | Xenocrates]], which Frescobot would fix when it

ran every month or two. However, the next time MathBot made its daily update, it would put the longer syntax back. This lasted from April 2010 to October 2012 and included 41 back-and-forth reverts between the two bots. This is a prime example of bot-bot task conflict in Hinds & Bailey's typology, even though there was no fundamental disagreement between the bot's developers about what ought to be done. The bots were still programmed with opposing tasks, and so they continued to revert each other. It did take over two years before it was noticed, but it was fixed by the bot's developer upon it being brought up in a talk page discussion.

*6.5.2 Mathbot's disambiguation links.*

In examining the Mathbot/FrescoBot conflict, we found a deeper case of conflict. Mathbot would also link directly to pages that were redirects to other articles, as well as linking to disambiguation pages. This curated list of all the articles, redirects, and disambiguation pages was used by members of WikiProject Mathematics as a helpful index to coordinate work. However, English Wikipedia guidelines discourage directly linking to redirects and disambiguation pages. There are several bots run by members of Wikiproject Disambiguation that will automatically fix redirects and links to disambiguation pages. These bots would "fix" the links to redirects, but the next time Mathbot updated the list, it would put the link to the redirect back.

When one of the Wikiproject Disambiguation members noticed this, they started a discussion on MathBot's developer's talk page. MathBot's developer was insistent that the index of links was useful and urged the developers of the redirect fixing bots to add an exception. They disagreed and argued that any lists used for Wikiproject Mathemathics shouldn't be in the main article namespace, but instead be pages in WikiProject Mathematics' namespace. The debate first moved to the WikiProject Mathematics talk page, then to a separate discussion and straw poll, which was more broadly advertised to get more third-party participants. This can be classified as a genuine conflict with dozens of participants. In Hinds & Bailey's typology, this conflict was first about the task, then the process, then it began to get interpersonal for some members. Over two months and 17,500 words of debates later, a consensus was reached that Mathbot would abide by the redirect and disambiguation guidelines in the main article lists, then curate separate lists in the Wikipedia namespace under WikiProject Mathematics.[16]

*6.5.3 Archiving links: AnomieBot vs CyberBot II.*

In July 2016, AnomieBOT and CyberBot II also had an 41 revert sequence on a single page over the course of only 4 days, which is the most intensive bot-bot revert war in our dataset in terms of reverts per page per day. AnomieBOT was approved to fix broken references, while CyberBot II was approved to add a link to the Internet Archive for references with dead links. However, CyberBot II had a bug that caused it to remove part of the reference instead of adding the Internet Archive link. AnomieBOT would detect the broken reference and would fix it, which CyberBot II would then remove. This bug affected CyberBot II's activity on 15 pages: this back-and-forth lasted for 41 reverts on the article about "Foreign relations of the Central African Republic", 35 reverts on the article about the songwriter Rico Love, 31 reverts on the article about the broadcaster Dougie Vipond, and more. In total, these bots reverted each other 396 times — one constantly adding links to Internet Archive pages and then the other removing them.

This can be considered a task conflict between the two bots in Hinds & Bailey's typology, although there was no conflict at all between the two bots' developers. Unlike the conflict over Mathbot, there was no fundamental disagreement about whether these references ought to appear or not that was fought through bots. Both bots were developed to cooperate with each other and their developers fully agreed with each other about what tasks ought to be done and how, but the bug caused CyberBot II to change its task from fixing references to removing them. When the issue was raised to CyberBot II's developer by multiple Wikipedians, the developer quickly disabled the bot, investigated, identified the bug, fixed it, re-enabled the bot, and wrote a short note explaining what went wrong. [17]

---

[16] http://tinyurl.com/mathbotpoll

## 7 COMMENT CODING

### 7.1 Overview

The previous section gave qualitative descriptions of different kinds of bot-bot revert events, giving enough of the local context to understand to what extent they are or are not conflict. On its own, these cases are sufficient counter-examples to critique the assumption that bot-bot reverts necessarily indicate conflict. These cases also gave us a set of different kinds of bot-bot revert events, which we had classified as various kinds of bot-bot conflict or non-conflict. However, we still did not have an answer to our empirical research question: *How many of the bot-bot reverts in our dataset are instances of genuine conflict like the one between AnomieBot and CyberBot II, and how many were cases of routine, even collaborative work like fixing double redirects?* As we had hundreds of thousands of revert events, manually reviewing and investigating them all ourselves was not feasible. Because the work requires substantial trace literacy of the Wikipedian community, crowdsourcing approaches would also be lacking.

We then realized we could make heavy use of the edit summaries (or edit comments) that any Wikipedian editor can leave for any edit they make. Edit summaries are an important and widespread way in which Wikipedians describe their work, as they are displayed to users who browse the revision history for a page, the recent changes feed, or lists of contributions by a single user. Geiger & Ribes's trace ethnography research [28, 29] discusses how Wikipedian vandal fighters leave notes, tags, and traces for others to follow in these edit summaries. Similarly, bot developers on the English Wikipedia are required to have their bots leave "informative messages" in the these edit summaries describing the task they are performing. This makes it somewhat easy to programmatically determine the kind of work that the bots are doing. However, these summaries are records of what the bot developer has programmed the bot to say the edit is doing (as opposed to what the bot is actually doing). In line with Star's discussion of data as veridical versus representational [73], these comments must be interpreted and aggregated with care. As in the case of AnomieBOT and CyberBot II (section 6.5.3), CyberBot II's bug caused it to remove references while still giving the edit summary "Rescuing 1 sources. #IABot".

### 7.2 Identifying patterns from edit summaries

We identified over 60 different patterns associated with different types of bot-bot revert activity, which each map onto one of the sections presented in detail in 6. We generated these patterns based in part on our prior experience in bot development and Wikipedia editing (we discuss our positionality in section 6.1). We found that many of the tags, markers, links, and abbreviations left in bot edit summaries were the same that Wikipedians use in discussing content-specific issues, as bots are delegated editorial work that humans would have to do manually.

In seeking to identify patterns that would match particular kinds of bot-bot revert events, we also used exploratory computational analyses of edit metadata and content (examining what was changed in each edit). One approach that provided much insight was grouping edits by reverting comment and counting the most frequently occurring comments, as well as counting the most commonly occurring words and n-grams in comments. When we found a new pattern, we would validate and refine it by spot checking a random sample of diffs. This was to help us be confident that the pattern captured the right kinds of edits and did not capture edits outside of the type.

---

[17] http://tinyurl.com/talkcyberbot

We also relied on quantitative metrics to help explore these data, such as the time to revert or sequences of reverts on a single page. We would focus on cases that had abnormally high or low values for these metrics, as well as sample from the median and mode. In some cases, we identified what appeared to be a genuine conflict between two bots, while in other cases, we identified broader kinds of non-conflict routine work in Wikipedia. Our GitHub repository[18] and our Figshare code & data repository [26] also contains the software code used to parse the dataset for each pattern, a table of sample patterns and matching comments, and a random sample of diffs for each bot type for each language version.

We combined pattern matching of edit summaries with matching of bot usernames to classify 6 of the cases that we identified as cases of conflicts (labeled botfight: then a descriptor). These include the Russbot / Cydebot edit wars over category renaming, the Mathbot mathlist updates, the Cyberbot II vs AnomieBot edit wars over date tagging, CommonsDelinker and ImageRemovalBot reverts about images in articles, and a revert of an approved bot that malfunctioned and was temporarily blocked. We found this last one exclusively in the edit comments, which had the text "Undoing massive unnecessary addition of infoboxneeded by a (now blocked) bot."

### 7.3 Validation

These edit summaries are based on the bot operator specifying what kind of task the bot is performing. In order to validate this, we generated diffs (showing what was added/removed in an edit) for a random sample of all reverted edits in each classified type per language: 1% of reverts for types with more than 10,000 cases and 100 reverts for smaller types. We have published these diffs on our GitHub repository for further spot checking and validation. These sample diffs were extremely useful in identifying further cases of bot-bot conflict, which helped us extend and refine our pattern matching. For example, one previous version for pattern matching in interwiki cleanups included searching for an interwiki link, e.g. [[:de:Deutschland]]. But we found that some bots would include interwiki links to their own language version for various reasons, to do work other than interwiki link cleanup, so we excluded interwiki links to the current language version.

## 8 ANALYSIS OF CLASSIFIED REVERTS

### 8.1 What proportion of bot-bot reverts are various types?

Once we had a dataset of bot-bot reverts classified by type and an understanding of what types could be taken to be bot-bot conflict vs non-conflict, we could learn what proportion of our dataset was made up of each type. Our findings illustrate that an overwhelming proportion of the bot-bot reverts are fixing double redirects and curating interwiki links. This occurs across language versions: table 2 shows the proportion of bot-bot reverts in each language that were classified in each type. In English Wikipedia articles, double redirect fixing accounted for 45% of all bot-bot reverts, interwiki link work for 49%, and 1.07% of cases were unclassified. The English Wikipedia bot fights we identified comprised 1.39% of bot-bot article reverts, although there are likely other cases we have not covered. Across all language versions, 3.5% of bot-bot reverts to articles were either not classified, classified as 'botfights', or contained 'revert' in the edit summary.

### 8.2 Time to revert plots by categorized type

Figure 6 plots a kernel density estimation (KDE) of time to reverts for each classified type. High values indicate high probability that a kind of revert will be made at that time period: for example, protection template cleanup spikes at 7 and 30 days, which are the two most common durations for page protection. This is because one bot will add a notice that a page is protected when an administrator protects it, than another bot will remove the notice when the protection expires. These plots also show how the most common types of bot-bot reverts (interwiki link cleanup, fixing double redirects, and other template work) predominantly take place months and years after the reverted edit was made. In contrast, edits that have "revert" in the edit comment (which strongly indicates conflict) as well as cases we have manually flagged as "botfights" have distributions that peak in the minute to day range.

---

[18] https://github.com/halfak/are-the-bots-really-fighting

| *type / language* | es | de | en | zh | ja | pt | fr |
|---|---|---|---|---|---|---|---|
| **category work** | 0.16% | 0.01% | 0.73% | — | — | — | 0.48% |
| **fixing double redirect** | 14.04% | 1.8% | 45.15% | 8.58% | 0.85% | 3.14% | 5.78% |
| **interwiki link cleanup method 1** | 23.56% | 30.17% | 34.2% | 34.07% | 14.5% | 22.66% | 17.95% |
| **interwiki link cleanup method 2** | 55.51% | 65.34% | 15.15% | 54.89% | 79.46% | 69.15% | 73.25% |
| **other w/ revert in comment** | 1.62% | 0.05% | 0.41% | 0.6% | 0.03% | 0.04% | 0.01% |
| **protection template cleanup** | — | 0.02% | 1.16% | — | — | — | — |
| **template work** | 0.2% | 0.01% | 0.52% | — | — | — | — |
| **other classified (not conflict)** | 0.21% | 0.02% | 0.23% | 0.01% | 0.02% | 0.06% | 0.71% |
| **identified botfight** | 0.02% | 0.03% | 1.39% | 0.01% | — | — | — |
| **not classified** | 4.69% | 2.56% | 1.07% | 1.85% | 5.14% | 4.94% | 1.82% |

Table 2. Proportion of bot-bot reverts classified in each type by language

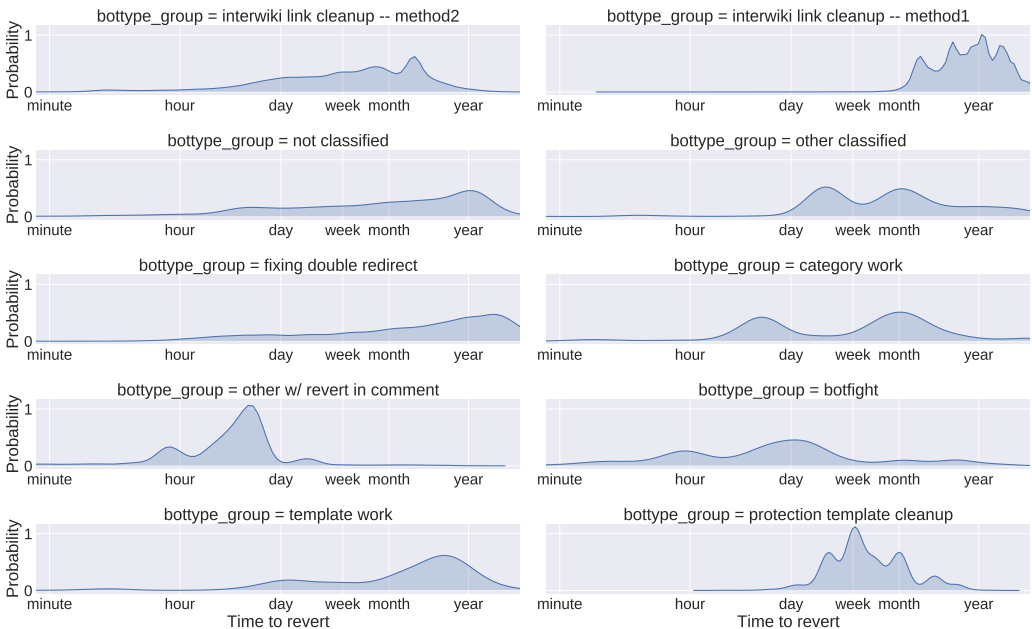

Fig. 6. Kernel Density Estimation showing the distribution of time to revert for different types of bot-bot reverts. Y axis for all plots is the KDE probability a revert takes place at that duration.

## 8.3 Suspected bot fights

We then filtered this dataset to examine the cases where there was more than one revert per bot pair on the same page, which indicates a reciprocal revert and is suggestive of potential conflict (see section 5.3). We then filtered it to exclude cases where the time to revert was less than 180 days, as such a long duration between edit and revert is indicative of non-conflict. Table 3 counts the number of reverts in this filtered set by language and the above categorization. As we explained in section 6, these types of bot-bot revert events ought not to be considered conflict for various reasons. There are then 962 unclassified bot-bot reverts that match these criteria for possible conflict, which is a tiny proportion of all bot-bot reverts across the seven languages over 15 years. In total across our 7 languages, identified botfights, cases with 'revert' in the comment, and non-classified cases in this filtered dataset number 3,007 reverts, which is approximately 0.5% of all bot-bot reverts to articles.

| type / language | de | en | es | fr | ja | pt | zh | total |
| --- | --- | --- | --- | --- | --- | --- | --- | --- |
| **category work** | 6 | 350 | 0 | 0 | 0 | 0 | 0 | 356 |
| **fixing double redirect** | 108 | 8925 | 690 | 86 | 20 | 113 | 446 | 10388 |
| **interwiki link cleanup method1** | 2 | 4 | 4 | 6 | 4 | 3 | 2 | 25 |
| **interwiki link cleanup method2** | 2906 | 2171 | 2284 | 2567 | 2059 | 2318 | 2169 | 16474 |
| **other w/ revert in comment** | 1 | 168 | 12 | 2 | 4 | 5 | 121 | 313 |
| **protection template cleanup** | 0 | 258 | 0 | 0 | 0 | 0 | 0 | 258 |
| **template work** | 0 | 33 | 0 | 0 | 0 | 0 | 0 | 33 |
| **other classified (not conflict)** | 1 | 97 | 0 | 1 | 0 | 0 | 2 | 101 |
| **identified botfight** | 10 | 1722 | 0 | 0 | 0 | 0 | 0 | 1732 |
| **not classified** | 86 | 355 | 155 | 59 | 145 | 95 | 67 | 962 |

Table 3. The number of bot-bot reverts (per language and classified type) where time to revert < 180 days and the revert was part of a case where there was more than one revert per bot pair on the same page

## 9 DISCUSSION
### 9.1 Bot-bot reverts on articles are not a major governance concern in Wikipedia

Our study has given crucial context to the phenomenon of bot-bot reverts in Wikipedia, showing they are generally far less of an issue than has been depicted in media coverage of the EGBF article. We found that the overwhelming majority of bot-bot reverts on articles are better characterized as routine, productive, and even collaborative work between bots. In our most conservative estimate, the proportion of bot-bot reverts to articles that are conflicts between bots — in the sense that differences in how the bots are programed lead to opposing bots contesting the content of articles (i.e. a bot-bot edit war) — is no more than 4%. With additional assumptions and methods, we can confidently estimate that approximately 0.5% of bot-bot reverts to articles constitute bot-bot conflict. We did find some genuine bot-bot conflicts, and it would be odd if we did not, given how many different bots have been created and deployed across Wikipedia's 15+ year history.

Hinds & Bailey's typology of task, process, and interpersonal conflict [38] proved to be quite useful in distinguishing between such cases, although we also needed to distinguish between bot-bot conflict (in which article content was contested by automated software agents) and conflict between human Wikipedians about bots, which may or may not co-occur with bot-bot conflict. We found bots with bugs that led them to task conflict with each other even though their developers had no conflict at all. In these cases (like AnomieBot and CyberBot II over article references), the developer of the misprogrammed bot fixed the issue soon after they were notified. We also found cases where two bot developers of approved bots entered into classic task conflict with each other, strongly disagreeing over what ought to be done in Wikipedia. This includes the case of Mathbot versus the various disambiguation and redirect fixing bots.

We also found bot-bot process conflict, where two bot developers agreed on the same fundamental goal, but had different encoded procedures about how to implement it that led to conflict, such as Russbot and Cydebot over article categorization. Finally, we found cases where conflict turned from task or process conflict into interpersonal conflict between bot developers, which happens all too often with editing Wikipedia articles as well. Yet to the best of our knowledge, there are no current, ongoing cases where bots are engaging in the kind of conflict depicted by the EGBF paper (which used a dataset from 2001-2010). In fact, quantitative metrics indicate that even this problematic operationalization of bot-bot conflict has been dramatically decreasing since 2013, as shown in section 5.

### 9.2 Focus on how conflict is resolved, rather than the presence of conflict

Drawing on classic CSCW literature about conflict and cooperation [15, 48], we do not believe that the mere presence of conflict between bots or even between Wikipedians about bots means that the Wikipedian community is somehow failing in the area of automation regulation. It is more useful and generative to ask what happens when such conflict arises, as successful and prompt conflict resolution is a sign of a healthy and developed organization. It is thus crucial to emphasize that out of all of the cases of bot-bot reverts we examined that did indicate some semblance of conflict, we were not able to find a single one where Wikipedians themselves had not already identified the issue and resolved it. In rare cases this took an extended period of time, but the kind of intense, back-and-forth edit wars over the content of articles was generally identified and fixed quickly.

These discussions between bot developers generally took place in a manner similar to discussions over Wikipedia article content. There is a set of policies which govern bot development, as well as a set of policies and guidelines which govern and guide participation by all users (bot and human). This governance environment is decentralized, but operates through increasingly escalating venues, which draw in a wider and wider population of Wikipedians [9, 19, 28, 50, 74]. We see this same governance structure at work in bot development, confirming previous case studies of Wikipedian bot governance: cases of conflict are rare but still take place [24, 62]. That said, Wikipedia talk pages in general can be a hostile environment, particularly for newcomers and members of underrepresented groups, and this has been the focus of recent work on Wikipedia's notoriously large gender gap among editors [52, 53, 59, 72].

### 9.3 ~~What~~ Who is a bot? The bot-developer assemblage as an analytical unit

At a broader level, one of the starkest differences between our study and the EGBF study is that for us, bots are inextricably linked to their developers, rather than autonomous agents who can be studied independently of the people who develop and operate them. Speaking to this, many Wikipedia bot developers name their bot not after the task it does, but after themselves: CydeBot is run by a user named Cyde, Addbot is run by Adam Shoreland (whose username is Addshore), CyberBot is run by a user named Cyberpower678, Xqbot is run by a user named Xqt, and more.

As many of our in-depth cases of bot-bot conflict and non-conflict illustrate, the actual programming of the bot to complete a particular task is only a part of what bot developers do. Like studies of open source software development have shown, there is also a substantial amount of social-organizational work that takes place behind the scenes [11, 17]. Wikipedian bot developers must speak and advocate for their bots in the Wikipedia community, which is a requirement of the bot policy. Bot developers make proposals to the Bot Approvals Group, respond to issues and concerns from BAG members and other Wikipedians before and after approval, debate and argue about to what extent existing policy supports a particular task, fix bugs when they appear, respond to feature requests, and interact with other bot developers — particularly when their bots are doing similar kinds of work. Bot developers who do not sufficiently respond to issues raised about approved bots are liable to having their bots blocked.

Literature in and around CSCW (especially STS and Organizational Studies) has long theorized agency as a heterogeneous, socio-technical assemblage that is co-constructed between humans and technology [6, 10, 66], and our study illustrates the continued importance of this approach, even in quantitative, computational methods. It is one thing to ask how often bots as automated software agents get into cycles of conflict with each other over the content of articles based on differences in their programmed directives. It is an entirely different question to ask about the dynamics of conflict when we see bots as assemblages of code and a human developer/advocate, who is responsible for operating the bot in alignment with Wikipedia's complex policy environment.

### 9.4 What about conflict between Wikipedians over what bots should do?

We are quite confident that bot conflict as defined by bots contesting the content of articles is relatively infrequent, short-lived, and causes little damage. However, there can be substantial, intense, and ongoing conflict between Wikipedians over what bots ought to be created. As our study was based on investigating bot-bot reverts, we have not generally discussed the conflicts that Wikipedians get into over opposing ideas about what bots should do when those conflicts do not lead to bot-bot reverts. This is the kind of epistemological issue we discussed in section 3, similar to how Tufekci [76] discusses the kinds of engagement with tweets that are not neatly encapsulated in a JSON field from the API. We did find many such cases in our research, which should be investigated in future work. There are only a handful of cases in the literature specifically focusing on bot controversies in Wikipedia [29, 62]. The thousands of bots that have been proposed and debated across Wikipedia provide a rich set of cases for the qualitative and quantitative study of how communities collectively decide how they want to use automation in their platforms.

### 9.5 Beyond pluralism: mixed-method collaboration and data contextualization

As we extensively discussed in section 3, one of the major contributions of this work is in presenting an alternative way of engaging with large-scale datasets made from records of the traces users leave behind in software platforms. Our approach took less of a veridical approach to our dataset of bot-bot reverts, as we instead saw such data as a starting point for further exploring cases and issues. Traditional narratives all too often position quantitative and qualitative methods in opposition to each other. Even the increasingly prevalent pluralistic position still frequently positions these approaches as orthogonal, even incommensurable ways of answering the same research question from different directions. Our mixed-methods approach is an integrative, iterative synthesis, not a coordinated pluralism: there is a column in a dataframe that would not exist without ethnography, as well as thick descriptions of cases that were found by sampling from the extreme ends of a statistical distribution.

We sought to present not only a well-researched answer to our empirical research question about the extent to which bot-bot reverts constitute bot-bot conflict, but also an example of how to cook data with care. We have given a substantial level of detail in our methods, processes, intentions, strategies, and positionalities — as well as made our full computational analysis pipeline public on GitHub[19] and Figshare [26] — in the hope that future researchers can learn from our experience. There are many specific lessons that Wikipedia researchers can learn from our paper, such our metrics, heuristics, strategies for distinguishing between different kinds of conflict. We also hope that researchers of all social computing platforms can benefit from how we sought to bring the goals of computational social science and qualitative contextual inquiry together.

---

[19] https://github.com/halfak/are-the-bots-really-fighting

### 9.6 Limitations and assumptions

*9.6.1 Limitations in found data.* As we discussed in section 3, our research — like that of a large and growing portion of research in CSCW and social computing — relies heavily on found data. Wikipedia's databases have a complex, messy 15+ year history, and their primary purpose is to host content for the Wikipedian community, not to facilitate social science research. Our main dataset is also based on bot-bot reverts, so there are many forms of conflicts around bots that do not present as bot-bot reverts (as we discussed in 9.4). As long-time Wikipedia researchers, we are aware of many limitations and issues with how the MediaWiki platform stores activity in its database, as well as interpretive ambiguities in the community. We have sought to conduct and limit our analyses accordingly. However, there are likely biases in how activity is recorded that we (and other Wikipedians) are unaware about. We do follow the various technical venues Wikipedians have for discussing these issues, which are of interest to a subset of volunteers and WMF staff.

*9.6.2 Limitations in trace ethnography.* Our paper relies on trace ethnography, and all ethnography requires some trust in the ethnographer's ability to have understood the culture and lifeworld in which they were embedded. Wikipedia is a far more public and archived space than most ethnographic fieldsites, and we have included various links throughout our descriptions that let readers follow these traces themselves. However, part of the point of trace ethnography is that these data do not speak for themselves; they must be learned as a criterion of membership in a community of practice. Having two researchers with extensive experience working in Wikipedia made it possible for us to double-check each other's interpretations and explications, and we spent much time collectively talking with each other about how we ought to make sense of this data. We also talked to Wikipedians about our interpretations of some of these cases, triple checking our assumptions.

For our comment coding in section 7, our pattern matching could be overly aggressive, capturing cases that are not in the intended type. We have done our own random spot checking, and our GitHub repository contains our pattern matching code and random samples of matching cases for further review. Finally, English is both of our primary languages and we have participated on English Wikipedia far more than other language versions. While we both have some working knowledge of various other languages and can partially rely on automated translation and Wikipedians' own cross-language translations, our ability to dive deep into non-English cases is limited. There are likely additional cases of bot-bot conflicts over article content in English and non-English Wikipedias that we have not identified in this article, which we leave for future work.

*9.6.3 Limitations in computational analysis.* Our analysis of bot-bot reverts is based on identity reverts, and it is possible that other approaches, like partial reverts, would identify additional cases with different proportions of conflict. Finally, we recognize that we are humans and are completely capable of making mistakes, errors, and bugs in our code, especially as our computational pipeline is complex and has many stages. While we have sought to follow best practices in software engineering and open & reproducible research (see [47]) to the best of our ability and experience, we must entertain the possibility of unintentional, undiscovered errors. We invite others to explore, re-run, and audit our data and code repositories, and to facilitate this, we have provided extensive documentation of our code and data so others can meaningfully understand our analysis pipeline.

## 10 CONCLUSION

In this paper, we have made several contributions to the study of both conflict and automation in CSCW systems. At a topic specific level, this study challenges the results of previous research. We present a more nuanced understanding of bot-bot reverts in Wikipedia that shows these events are generally not a cause for concern in the way mass media coverage has imagined. We have given in-depth, thick descriptions of bot work in Wikipedia's particular socio-technical ecosystem that contextualizes two of the largest kinds of bot-bot reverts as routine, collaborative work: double revert fixing and interwiki link curation. We identified cases of genuine conflict, then discussed how and why they arise, as well as what steps were taken to resolve them. This paper moves the conversation beyond identifying bot-bot reverts and towards a more holistic understanding of bot governance, as we distinguish between bot-bot conflict (where bots contest the content of articles based on opposing programmed directives) and conflict between Wikipedians about bots. We also presented a new synthetic approach to identifying Wikipedia bot accounts automatically across language versions, and have released a new dataset of bots that will support future research.

At a concept level, this paper advances our understanding of both conflict and automation in CSCW. We have extended longstanding scholarship that has theorized different kinds of conflict in organizations, as well as extended scholarship that theorizes the roles that different types of conflict can play in social worlds. We have shown that approaches and metrics appropriate for studying human activity are not necessarily translatable to bot activity. Scholars studying a wide variety of social computing platforms should use a diversity of strategies to verify their methods when studying automated agents. We have shown that Wikipedia's bot regulatory system does not generally lead to the kinds of deeply problematic bot-bot conflicts that would be sufficient cause to question the wisdom of Wikipedia's decentralized, consensus-based governance model in other platforms or areas. We have also extended the theoretical literature on automation and agency in CSCW, as our findings further speak to the need to study automated agents and their developers/operators as a single unit, rather than in isolation.

Finally, at an epistemological level, our paper makes contributions to the broad, cross-disciplinary conversations about how social science is changing given the rise of found data collected by social computing platforms that log user activity. We have shown the benefit of a trace ethnographic approach, in which seek to understand data in the context it has within a community of practice, based on traditional ethnographic methods of participant-observation, interviews, and archival work. Our study also illustrates how trace ethnography can scale down to the level of specific thick descriptions as well as up to the level of aggregate statistical metrics. We have also shown the benefits of seeing this kind of trace data as less veridical and more representational, taking data to be a starting point for further contextualization and interpretation, rather than the output of a sensor that more or less faithfully captures activity in the world. And at the broadest level, we have shown that these epistemological lessons (largely made from qualitative researchers) are also highly relevant for conducting quantitative, computational research. Our integrative, iterative approach blending computational social science and qualitative contextual inquiry provides a better understanding of the world than either method can alone (or even in sequence).

## ACKNOWLEDGEMENTS AND FUNDING

We would like to thank the anonymous referees for their valuable comments and helpful suggestions. We also thank various Wikipedian volunteers and Wikimedia Foundation staff for speaking to us about these issues, in particular Tilman Bayer and Adam Shoreland, the maintainer of Addbot. We also thank many people for their feedback and advice on this project, as well as for help and support on doing this kind of scholarship in an open and reproducible manner, including: Aaron Culich, Anissa Tanweer, Anne Jonas, Brittany Fiore-Gartland, Charlotte Cabasse-Mazel, Chris Holdgraf, Fernando Perez, Karthik Ram, Laura Nelson, Laura Norén, Lauren Ponisio, Matthias Bussonnier, M Pacer, Nelle Varoquaux, Nick Adams, Nick Merill, Noura Howell, Philip Stark, Richmond Wong, Yuvi Panda, and the Reproducibility & Open Science working group at the Berkeley Institute for Data

Science. We would especially like to thank Yuvi Panda for helping us set up and use a reproducible development environment for this project, as well as Yuvi's efforts to make Wikipedia's databases publicly queryable through projects like Quarry, which we also used in this project — and found to be crucial in making part of our data collection pipeline open and reproducible.[20]

This research is funded in part by the Gordon and Betty Moore Foundation (grant GBMF3834) and the Alfred P. Sloan Foundation (grant 2013-10-27), as part of the Moore-Sloan Data Science Environments (MSDSE)[21] grant to the University of California, Berkeley. Geiger is currently an ethnographer and post-doctoral scholar at the MSDSE-funded UC-Berkeley Institute for Data Science. Halfaker is currently a principal research scientist at the Wikimedia Foundation, which has not provided direct financial support for this specific work, although Halfaker has contributed to this work as part of his research staff position. The Wikimedia Foundation has also indirectly supported this work by maintaining a public, general purpose research infrastructure called Toolforge[22] (formerly called Tool Labs), which we also used.

---

[20] https://quarry.wmflabs.org
[21] http://msdse.org/
[22] https://wikitech.wikimedia.org/wiki/Help:Toolforge

# Operationalizing Conflict & Coordination Between Automated Software Agents   49:33